\documentclass[aip,rsi,floatfix,showpacs,superscriptaddress]{revtex4-1}
\usepackage{graphicx}
\usepackage{amssymb}
\usepackage{amsmath}
\usepackage{soul}
\usepackage[usenames]{color}
\usepackage{ulem}

\ifx\pdftexversion\undefined
\usepackage[dvips]{hyperref}
\else
\usepackage{hyperref}
\fi

\begin{document}

\title{Influence of Gold Coating and Interplate Voltage on the Performance of Chevron Micro-Channel Plates for the Time and Space Resolved Single Particle Detection}
\author{A. L. Hoendervanger}
\email[Electronic address: ]{Lynn.Hoendervanger@institutoptique.fr}
\affiliation{Laboratoire Charles Fabry, Institut d'Optique, CNRS, Univ. Paris Sud, 2 Avenue Augustin Fresnel 91127 PALAISEAU cedex, France}
\author{D. Cl\'ement}
\affiliation{Laboratoire Charles Fabry, Institut d'Optique, CNRS, Univ. Paris Sud, 2 Avenue Augustin Fresnel 91127 PALAISEAU cedex, France}
\author{A. Aspect}
\affiliation{Laboratoire Charles Fabry, Institut d'Optique, CNRS, Univ. Paris Sud, 2 Avenue Augustin Fresnel 91127 PALAISEAU cedex, France}
\author{C. I. Westbrook}
\affiliation{Laboratoire Charles Fabry, Institut d'Optique, CNRS, Univ. Paris Sud, 2 Avenue Augustin Fresnel 91127 PALAISEAU cedex, France}
\author{D. Dowek}
\affiliation{Institut des Sciences Moleculaires d'Orsay, UMR 8214, Univ. Paris-Sud, Bat. 350, 91405 ORSAY cedex, France}
\author{Y. J. Picard}
\affiliation{Institut des Sciences Moleculaires d'Orsay, UMR 8214, Univ. Paris-Sud, Bat. 350, 91405 ORSAY cedex, France}
\author{D. Boiron}
\affiliation{Laboratoire Charles Fabry, Institut d'Optique, CNRS, Univ. Paris Sud, 2 Avenue Augustin Fresnel 91127 PALAISEAU cedex, France}

\begin{abstract}
We present a study of two different sets of Micro-Channel Plates used for time and space resolved single particle detection. We investigate the effects of the gold coating and that of introducing an interplate voltage between the spatially separated plates. We find that the gold coating increases the count rate of the detector and the pulse amplitude as previously reported for non-spatially resolved setups. The interplate voltage also increases count rates. In addition, we find that a non-zero interplate voltage improves the spatial accuracy in determining the arrival position of incoming single particles (by $\sim20 \%$) while the gold coating has a negative effect (by $\sim30 \%$).
\end{abstract}

\maketitle

Micro-Channel Plates (MCP) detectors are  widely used in atomic and nuclear physics for the detection of charged particles and photons \cite{examples,Barat2000}.
They are also used to detect metastable rare gas atoms \cite{Vassen12}
because the internal energy of such atoms is sufficient to produce electrons on the MCP surface.
Combining these detectors with a crossed delay line anode allows one to reconstruct the full three-dimensional distribution of single particles falling with a well defined velocity onto the surface of the MCP \cite{Jagutzki02}.
In particular second ($g_{2}$) and third ($g_{3}$) order correlation functions
along the three spatial axes have been reconstructed,
giving access to information related to the quantum statistics and coherence of a cold atom cloud \cite{Schellekens05,Jeltes07,Hodgman11}.
In these experiments, the quantum efficiency of the MCP-Delay Line detector has been estimated to be of order 10\% \cite{Jaskula10}, a rather disappointing figure which needs to be improved if one wishes to investigate quantum correlations at the two- and many-body level \cite{Altman04, Folling05, Greiner05, Rom06, Guarrera08, Kitagawa11}.
The spatial accuracy of the MCPs-Delay Line ensemble used in \cite{Jaskula10} is typically $\sim 150~$$\mu$m, and the above mentioned experiments could benefit from an improvement in this figure as well. Efficiency of detection and spatial accuracy are crucial quantities in the context of ion studies as well and they have been the object of recent work \cite{Lienard05}.

In this paper we will describe a study of potential ways to improve these performances. Testing different MCP-Delay Line detectors with metastable atoms directly is impractical due to the vacuum requirements set by the use of ultra-cold gases, and therefore we will attempt to simulate the response of the MCP to metastable atoms by using ultraviolet photons. The impact of a UV photon will in general only release one starting electron, as the impact of He* does, whereas for example ions might release more than one. As discussed later, we consider that, since similar and small (of the order of 10~eV) energy scales are involved, what we learn about MCP performance with UV photons carries over to metastable rare gas atoms. In addition of course, our results may be of use to workers interested in detecting photons with MCPs.

\begin{figure}[h]
\includegraphics[width=.5\columnwidth]{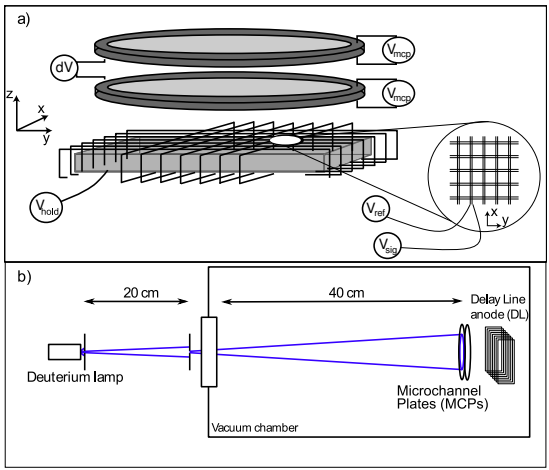}
\caption{\textsl{Experimental setup. {\bf a)} Details of the micro-channel plate mountings. Each plate is provided with a variable voltage $V_{{\rm mcp}}$. The output face of the second plate is set to 0~V. A variable voltage $dV$ is applied between the two MCPs. Finally, the delay line anodes are supplied with fixed voltages $V_{{\rm sig}} = 300$ V and $V_{{\rm ref}} = 250$ V for the transmission lines and $V_{{\rm hold}} = 150$ V for the anode holder.  {\bf b)} A deuterium lamp emitting photons in the range $185$ nm to $400$ nm is shone onto the MCPs. The light passes through two pinholes which control the flux. Inside the vacuum chamber the photons hit the MCPs.}}\label{Fig1}
\end{figure}

Several schemes have already been tested to improve MCP performance.
Adding a special coating, for example gold, aluminium or CsI, on the input face of the MCP has been found to increase the efficiency and the maximum count rate \cite{Fraser91,Tremsin1996} for photon detection.
For specific applications which rely on pulses of larger amplitude, another scheme consists of spacing the two MCPs, unlike the common chevron stacked configuration, and adding a voltage between the two slightly separated MCPs (see Fig.~\ref{Fig1}a). With such spaced MCPs, the additional voltage, which increases electron energy and influences their collimation in between the plates, should raise the efficiency and increase the resolution of the detector \cite{Rogers82,Firmani82}, but not all workers have observed such improvements \cite{Wiza77}.

Building on these ideas, we extend previous experimental investigations to the study of pairs of slightly separated MCPs using delay line anodes.
We investigate both the effect of the voltage difference between the plates and the presence of a gold coating by looking at overall count rates as well as at individual electronic pulses. We discuss in detail the consequences of the different configurations on the accuracy of these detectors in the particle counting mode. We conclude that the addition of a small voltage between the two plates can slightly enhance the detection efficiency and accuracy of such a detector.

%%%%%%%%%%%%%%%%%%%%%%%%%%%%%%%%%%%%
\section{General considerations}
%%%%%%%%%%%%%%%%%%%%%%%%%%%%%%%%%%%%%%
\label{Section:GenConsideration}

\subsection{Micro-channel plates (MCP)}

MCPs consist of many single channels which are basically electron multipliers. To operate they need a voltage $V_{{\rm mcp}}$ between both edges of each tube (\textit{i.e.} between both faces of the MCP), on the order of a kilovolt. A metal coating is therefore always evaporated on both faces of the MCP to form electrodes, such coatings being commonly mainly made from Nickel (as in Nichrome or Inconel coatings). When the voltage $V_{{\rm mcp}}$ is not high enough, there will be insufficient generation of secondary electrons when a particle hits the MCP. Raising $V_{{\rm mcp}}$ increases the ability of the MCP to start a cascade when an incident particle hits. At low $V_{{\rm mcp}}$ the gain increases exponentially with the voltage $V_{{\rm mcp}}$. For large $V_{{\rm mcp}}$ the gain of the detector saturates and increases linearly with $V_{{\rm mcp}}$. This happens when the presence of too many charges inside the channels screens the voltage $V_{{\rm mcp}}$. This effect is independent of the flux of incoming particles. At a fixed discrimination level used to distinguish signal from noise, the measured count rate will saturate with $V_{{\rm mcp}}$ since the height of the pulses increases with $V_{{\rm mcp}}$ and at a certain point all of the pulses will be above the discrimination level. This picture is valid at low flux.

When the flux of incoming particles is such that the arrival time between two incident particles is shorter than the time required to reestablish the electric field after the channel has been depleted, the second particle produces a smaller pulse.
Thus, whatever the gain of the detector (even when saturated), if the flux is too high, the detector will show "flux saturation". The flux at which this happens is dependent on $V_{{\rm mcp}}$. Adding a gold coating on the input plate can modify this behavior: the gold coated surface is more conductive than usual Ni-based coatings, thereby allowing faster replenishment of the channels and the detection of higher flux of incoming particles \cite{Fraser91,Tremsin1996}.

As a function of the flux of incoming particles, one can therefore identify three regimes at large $V_{{\rm mcp}}$: {\it (i)} at very low flux the number of counts increases linearly with the number of incoming particles with a slope defined by the detection efficiency of the detector; {\it (ii)} at moderate flux, the count rate increases more slowly than the number of incoming particles due to the flux saturation mentioned above; {\it (iii)} at large flux, the count rate hardly evolves with the number of incoming particles and is dependent on $V_{{\rm mcp}}$.

\subsection{Delay line anodes}

Delay lines are transmission lines that collect the electron pulses coming from the MCP. The transmission line is wound into a helical propagation line so the speed of the pulse in the relevant direction ($x$ in Fig.~\ref{Fig2}) is much smaller than the speed of the pulse in the wire itself (see appendix~\ref{AppendixC}). Dispersion ensures that the pulses created by a single electron shower in adjacent loops of the delay line overlap before reaching the end of the line \cite{Jagutzki02} (see Fig.~\ref{Fig2}). This leads to a single broadened pulse at the output whose center can be precisely determined  to a precision greater than the spacing between adjacent loops. Adjacent loops are separated by distances on the order of a mm and a propagation time of a few ns. Since the delay lines are built in such a way that dispersion combines the separate pulses on different loops into one larger one, pulse widths are typically a few ns. There is a delay line wound in both the $x$- and $y$-direction and from the arrival time signals obtained from both of these delay lines a spatial reconstruction can be made, and combined with the information of the arrival time on the MCP, one can reconstruct a 3D image.

\begin{figure}[h]
\includegraphics[width=.5\columnwidth]{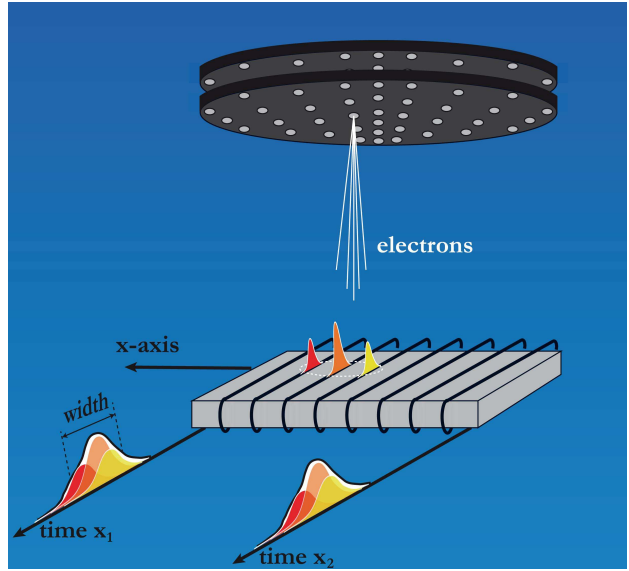}
\caption{\textsl{An electron shower exiting the second plate of the MCP pair is accelerated towards the delay line anode. During the time-of-flight the electron shower expands transversally and finally excites several loops of the delay lines. After propagation in the transmission lines, the pulses from the different loops overlap due to dispersion. A single broadened pulse is observed at the output of the delay lines.}}\label{Fig2}
\end{figure}

\subsection{Influence of interplate voltage $dV$}

Let us now consider the situation of Fig.~\ref{Fig1}a) with spaced MCPs when a voltage difference $dV$ is applied between the plates. We briefly comment on the electron dynamics in the interplate region \cite{Rogers82} and during the expansion from the second MCP to the delay lines. The voltage between the plates accelerates the electrons between the two plates which otherwise travel in a zero electric field free space, with the result of \textit{(i)} collimating the electron shower and \textit{(ii)} transferring energy to the electrons. As a consequence of having a larger energy, the electrons hitting the second plate have a higher chance of inducing a cascade in the channel. The increase of the voltage $dV$ should therefore have a positive effect on the efficiency of the detector and on the signal-to-noise ratio of the electronic pulses.

From simple arguments as put forward in \cite{Rogers82} we can attempt to draw a picture of the behavior of the electron cloud during expansion between the second MCP and the delay line anode. At fixed voltage difference between the bottom plate of the second MCP and the delay line (as in our experimental configuration), the number of loops where a pulse is produced by an incoming electron shower is mainly the result of two quantities: the initial size of the electron shower on the lower surface of the second MCP and the transverse expansion of the shower during the time of flight towards the delay lines. On the one hand, the smaller the number of micro channels excited in the second MCP ({\it i. e.} the smaller the initial size of the electron shower out of the MCP surface), the smaller the final size of the electron shower hitting the delay line. On the other hand, the faster the electrons exiting the second plate, the greater their transverse velocity and the larger the size of the electron shower hitting the delay line (see Appendix~\ref{AppendixA}). Adding a non-zero voltage difference $dV$ between the two plates clearly affects the size of the electron shower out of the second MCP (collimation of the interplate electron shower). It might also modify the energy of the electrons exiting the second plate. The net effect of $dV$ is therefore non trivial and it is worth being investigated.

%%%%%%%%%%%%%%%%%%%%%%%%%%%%%%%%%%%%
\section{Experimental Setup}\label{expsetup}
We study two different pairs of MCPs. Both pairs are new, Nichrome MCPs from Photonis USA, Inc. with a diameter of $1.245$ inch and were matched by the manufacturer. The micro-channels have a diameter of $10~\mu$m, a center to center spacing of $12~\mu$m and a bias angle of $12 ^{\circ}$. All technical specifications are given by the manufacturer and we have found no discrepancies in the resistances of the plates with the ones provided. One pair has a layer of $10~\mu$m of gold deposited on the input surface of the first plate. In this work we have not considered the alternative mounting consisting in using the gold surface as the input one of the second plate. In the following we will refer to the non-coated MCP pair when using two Nichrome coated MCPs and to the coated pair when using the pair with the gold coating.

The setup of the experiment is shown in Fig.~\ref{Fig1}. The MPCs are placed in a vacuum chamber with a pressure of $~10^{-6}$ mBar. Our mount has a non-conducting circular spacer with conducting surfaces between the two plates allowing us to apply an adjustable voltage difference $dV$ (see Fig.~\ref{Fig1}a). In our studies there is always a spacing between the two plates, whereas in the usual chevron configuration two MCPs are mounted directly on top of each other.

The voltage between the two MCPs ($dV$) can be varied from $0~$V to $250~$V. Below the MCPs is a delay line (Roentdek DLD80). The use of a delay line allows us to make a three-dimensional reconstruction of the signal gathered by the MCPs \cite{Jagutzki02}. In all the results of this paper, the voltage supplies of the delay lines are fixed  [see Fig.~\ref{Fig1} b)]: $V_{{\rm sig}}= 300~$V, $V_{{\rm ref}}= 250~$V, $V_{{\rm hold}}= 150~$V. The pulses from the delay line are decoupled from the reference voltage by a Roentdek FT12-TP feedtrough and decoupler which allows us to look at the pulse which is generated when a photon emitted by the source hits the MCP and triggers an electron shower. The signals are then amplified and discriminated by a Roentdek ATR-19 Constant Fraction Discriminator/Amplifier. The settings of the electronics are fixed throughout the article with the CFD level threshold set at $5~$mV. The amplified pulses are then processed by a Time to Digital Converter (CTNM4, IPN Orsay) with a coding step of $250~$ps. For counting the pulses from the delay line we take the pulses after the CFD/Amplifier and count them with a counter.

To simulate the metastable helium atoms (He*) we use ultra-violet (UV) light source, a convenient solution since UV photons share neutral charge and similar energy as He* (the latter being 19.8 eV). Therefore the impact of a UV photon will in general only release one starting electron, as the impact He* does, whereas for example ions might release more than one. The lamp is a Hamamatsu L6301-50 deuterium lamp with a spectrum ranging from $185~$nm to $450~$nm. The window through which UV photons enter the vacuum chamber is made of magnesium fluoride, which has a cut-off wavelength at $100~$nm. To control the incoming flux of photons on the detector we use two small pinholes, one right after the lamp and another one in front of the window of the vacuum chamber (See Fig.~\ref{Fig1}b). The diameter of the first pinhole ranges from $0.5~$mm to $2~$mm and the second pinhole has a fixed diameter of $1~$mm. We obtained a figure for the photon flux at different pinhole sizes by measuring the intensity of the lamp at $410~$nm with the use of an interference filter. We then used the spectrum provided by the manufacturer to obtain absolute values for photon flux for all wavelengths.

The lamp spectrum is broad, however only a small fraction of the photons are expected to contribute noticeably to the signal. This is due to the detection efficiency of channel electron multipliers in the UV range which decreases from $10^{-5}$ at $180~$nm to $10^{-9}$ at $250~$nm \cite{Paresce75}. As a matter of fact, we have operated with a Kodial glass window, with a high-pass cut-off wavelength of $\sim 250~$nm, resulting in an almost complete suppression of the number of counts on the detector. Therefore we are confident that we mostly detect the highest energy photons (close to $200~$nm) with energies of about $6-7~$eV.

%%%%%%%%%%%%%%%%%%%%%%%%%%%%%%%%%%%%%%
\section{Experimental Results}
%%%%%%%%%%%%%%%%%%%%%%%%%%%%%%%%%%%%%%
\subsection{Count Rate Versus Flux}
\begin{figure}[h]
\includegraphics[width=.5\columnwidth]{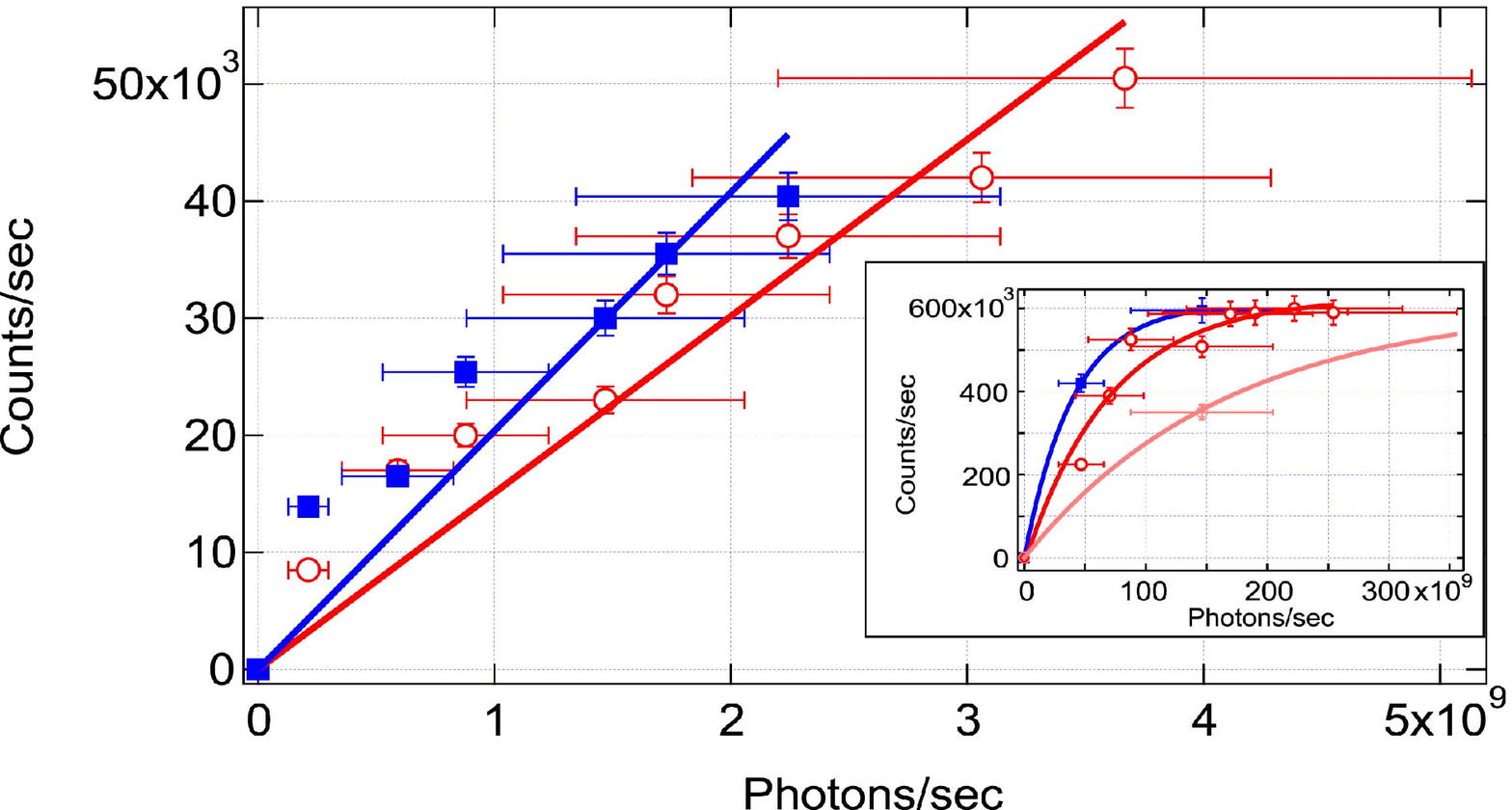}
\caption{\textsl{The number of counts with increasing flux. Red open circles: MCP pair without coating; blue filled squares: MCP pair with gold coating. Both pairs: $V_{{\rm mcp}}= 1000~$V, $dV = 0~$V. Lines are linear fits crossing the origin.
\\Inset: the count rate at higher flux. The count rate for the gold coated pair (blue squares) hardly changes with $dV$ (plotted is $V_{{\rm mcp}}= 1000~$V, $dV = 25~$V). $dV$ has a strong influence on the non-coated pair: Light red diamond $V_{{\rm mcp}}= 1000~$V, $dV = 0~$V / Red circles $V_{{\rm mcp}}= 1000~$V, $dV = 25~$V. Lines are exponential fits.}}  \label{Fig3}
\end{figure}

To study the characteristics of the two different pairs of MCPs, we first monitor the total number of counts from the MCPs as a function of the incoming photon flux, with the results shown in Fig.~\ref{Fig3}. The plotted number of photons hitting the active area of the MCP per second corresponds to the number of photons emitted by the lamp in a spectral range of $\sim 1~$nm at the wavelength $200~$nm according to our calibration. We note that this photon number slightly underestimates the total number of photons since the lamp emits a broad spectrum. For low flux (see Fig.~\ref{Fig3}) the count rate of the detectors increases linearly with the light intensity. The slope at low flux is $2.0 \times 10^{-5} $ counts/photon for the gold coated MCP and $1.5 \times 10^{-5}$ counts/photon for the MCP without coating. These values are comparable to those in Ref.~\cite{Paresce75} for photons with wavelengths of $ \leq 200~$nm.

For larger (intermediate) flux (see inset Fig.~\ref{Fig3}) the response deviates from the linear dependence, and starts to show flux saturation. In this situation the gold coated MCPs still have a better efficiency than the non-coated ones, with count rates larger by more than a factor $2$, even in the presence of a non-zero $dV$ on the plates.
We relate this observation to the lower surface resistance of the gold coated plates with respect to the non-coated ones, allowing a faster replenishment of the input plate \cite{Tremsin1996}. In addition we note that the work functions for gold ($5.1~$eV) and nickel ($4.6~$eV) are similar, implying that a similar quantum efficiency on the input plate (probability for an incoming photon to excite a first electron) is expected for both cases and would not explain our observation.
When the flux is even larger, the count rate saturates. The value of the saturation level is set by the properties of the electron channels, which we assume to be identical for both sets of plates as observed in Fig.~\ref{Fig3}.

%%%%%%%%%%%%%%%%%%%%%%%%%%%%%%%%%%%%%%
\subsection{Count rates vs $dV$ and $V_{{\rm mcp}}$}
\begin{figure}[t]
\includegraphics[width=.5\columnwidth]{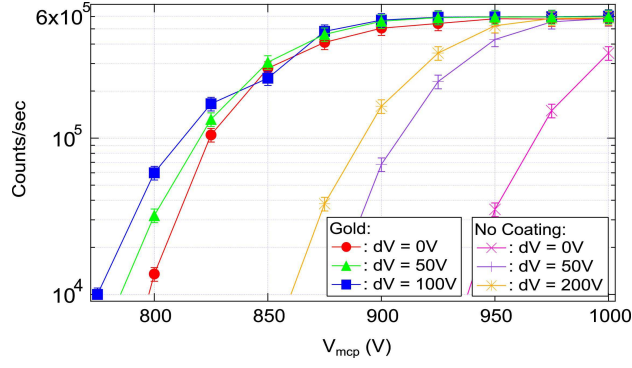}
\caption{\textsl{%%\textbf{Left:}
Count rates as a function of $V_{{\rm mcp}}$ for the two pairs of MCPs and for different voltage differences $dV$. Incoming flux $\sim 1.5 \times 10^{11}$ photons/sec.
}}\label{Fig4}
\end{figure}

Further characterization of the MCPs is done by investigating the response of the count rate to the gradual increase of $V_{{\rm mcp}}$.
In Fig.~\ref{Fig4} the increase in the number of counts with the total voltage over the separate MCPs ($V_{{\rm mcp}}$) at a given photon flux of $\sim 1.5 \times 10^{11}$ photons/sec is shown. We observe that the gold-coated detector exhibits gain saturation at a much lower total voltage ($V_{{\rm mcp}}$).

\begin{figure}[h]
\includegraphics[width=.5\columnwidth]{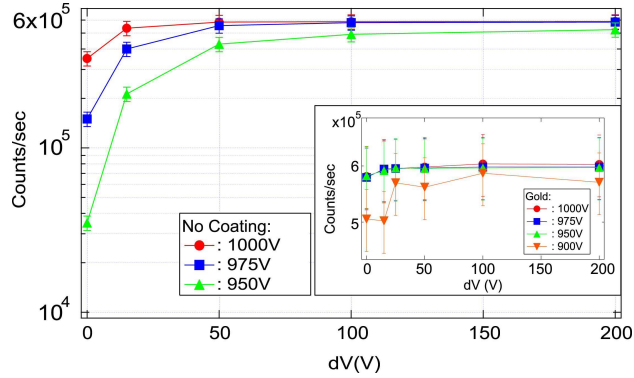}
\caption{\textsl{Count rates measured with the non-coated MCP pair as a function of $dV$. Inset: count rates measured with the gold coated MCP pair as a function of $dV$. Incoming flux $\sim 1.5 \times 10^{11}$ photons/sec.}}\label{Fig5}
\end{figure}

We now turn to a systematic investigation of the influence of the interplate voltage $dV$. The influence of $dV$ on the count rate of the MCPs is shown in Fig.~\ref{Fig5}. For the non-coated pair adding a voltage between the plates strongly increases the count rate of the MCPs. The effect saturates at $dV \simeq 50~$V. For the gold coated MCP, the count rate is already saturated and $dV$ has no effect. The observed increase in count rate with $dV$ differs with the findings by Wiza {\it et al} \cite{Wiza77} where the gain of the detector gets worse when there is a voltage between the plates. However, as demonstrated by Rogers and co-workers \cite{Rogers82}, the distance between the plates is crucial to the effect of $dV$. For instance when the plates are too close together the voltage may have a negative effect on the gain. Indeed we observed that tightening the MCPs too close to each other results in a drop of the count rate by a factor $5-10$. By contrast the results presented here are obtained in a favorable situation with a larger distance between the plates of about $\simeq 0.2~$mm. We have shown that in these circumstances one can compensate with the voltage $dV$ for a lack of gold coating and for a lower $V_{{\rm mcp}}$ with respect to the count rate.

%%%%%%%%%%%%%%%%%%%%%%%%%%%%%%%%%%%%%%
\subsection{Pulse shapes}

The three-dimensional reconstruction of particles falling onto the MCPs is possible thanks to recording the four arrival times of the pulses ($t_{x1}$, $t_{x2}$, $t_{y1}$ and $t_{y2}$) out of the two delay lines $x$ and $y$ \cite{Jagutzki02}. The position of an incoming particle is extracted from the quantities $t_{x1}-t_{x2}$ and $t_{y1}-t_{y2}$. Since the position of the particle is determined from the timing signals, the ability of the detector to correctly extract this position can depend on the amplitude and width of the pulses recorded from the delay lines. We thus pay attention to those quantities, in particular to understand in further details the influence of $dV$ on the timing signals.

%%%%%%%%%%%%%%%%%%%%%%%%%%%%%%%%%%%%%%
\subsubsection{Amplitude}
\begin{figure}[h]
\includegraphics[width=.5\columnwidth]{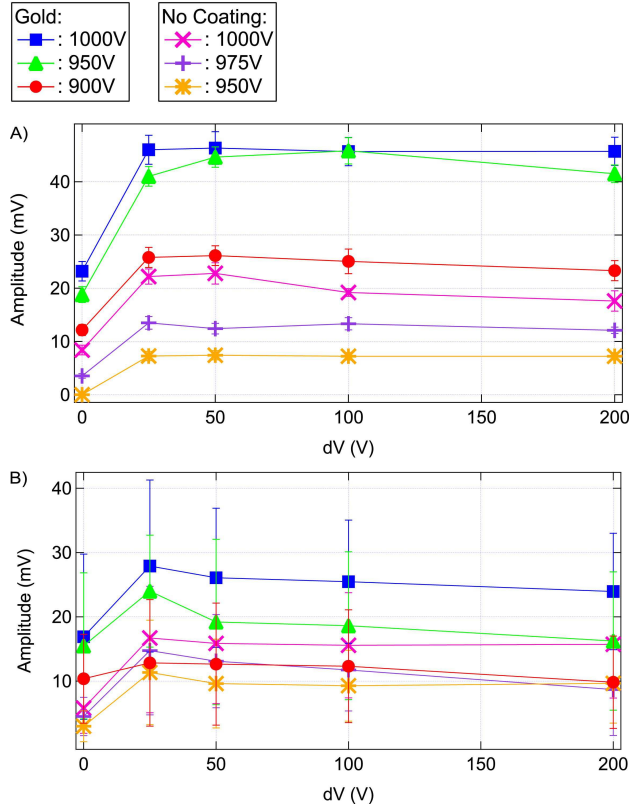}
\caption{\textsl{{\bf A)}The amplitude of the pulse from the input plate, for non-coated and gold-coated micro channel plates. {\bf B)}  The amplitude of the pulses from the delay line, averaged over all four channels.
}}\label{Fig6}
\end{figure}

We plot the amplitude measurements in Fig.~\ref{Fig6} A) and B). First we note that in the absence of voltage between the plates ($dV=0~$V), the pulse amplitude on the input plate is larger with the gold coating, which is consistent with the higher count rates observed for the gold coated plate. The effect on the pulse amplitude is observable on the delay line signal as well. As a consequence and for a fixed level of the CFD, the number of counts is larger for the coated plate (see Fig.~\ref{Fig4}). Secondly, we clearly observe an effect of $dV$ on the pulse amplitude, as the electrons gain energy by being accelerated between the plates. We see that the amplitude of the pulses from the front increases as $dV$ increases and then saturates, whereas the pulses from the delay line have a maximum at $\sim 25~$mV. As anticipated in Section~\ref{Section:GenConsideration}, we believe that in the delay lines measurement two competing effects contribute: faster electrons have a higher probability to fire a channel of the second MCP but, then, they also extend over a larger area during the expansion time to the delay line. We will discuss this interplay in more detail below.

%%%%%%%%%%%%%%%%%%%%%%%%%%%%%%%%%%%%%%
\subsubsection{Width}
\begin{figure}[h]
\includegraphics[width=.5\columnwidth]{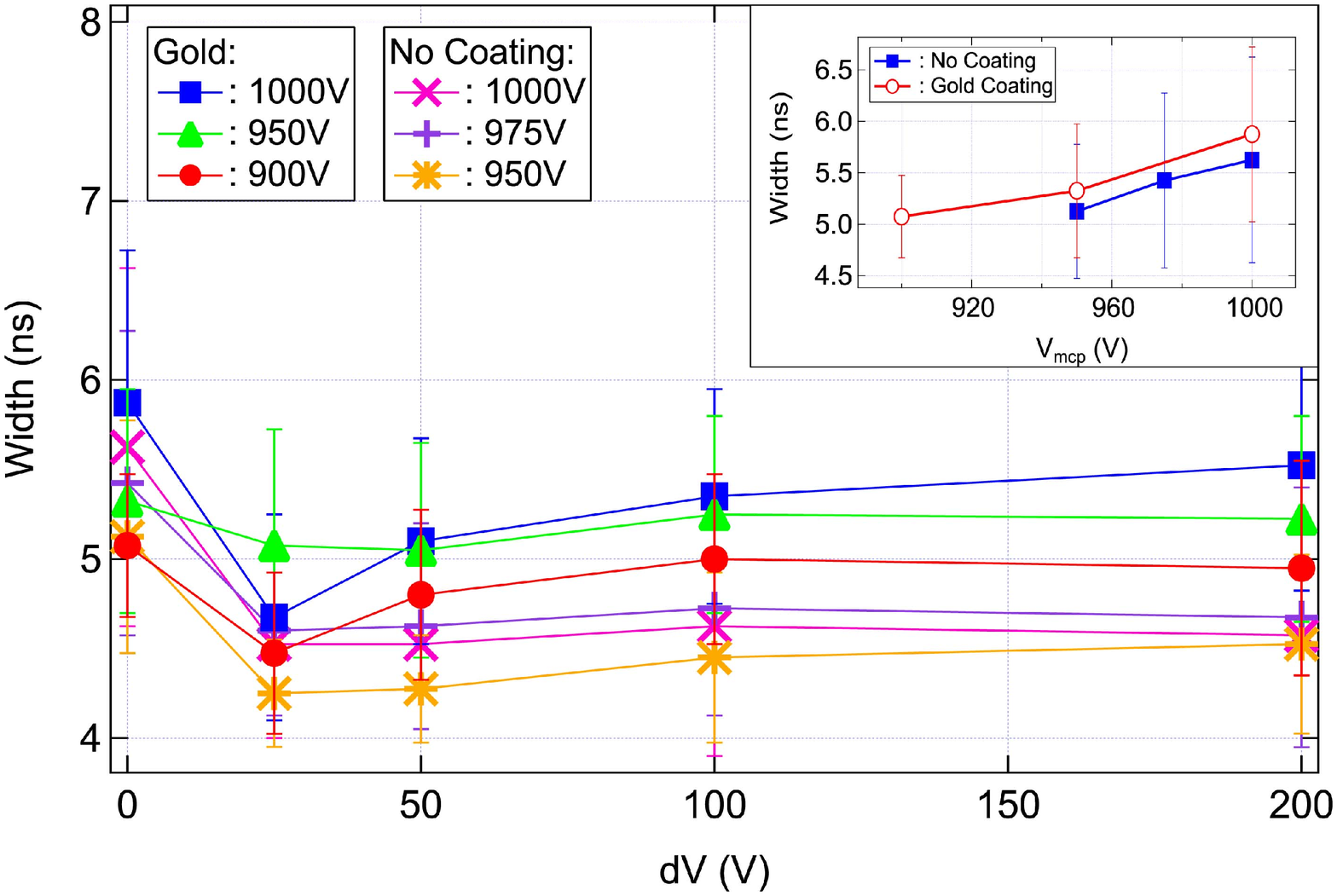}
\caption{\textsl{The width of the pulses from the delay line, averaged over all four channels. Inset: The width of the pulses at $dV=0~$V for increasing $V_{{\rm mcp}}$.}}\label{Fig7}
\end{figure}

The recorded widths of the timing signals are plotted in Fig.~\ref{Fig7} for different experimental situations. As predicted above, the typical width of the pulses is of the order of $5~$ns, due to dispersion in the transmission lines (see Section \ref{Section:GenConsideration}). At $dV=0~$V (see inset Fig.~\ref{Fig7}) the width of the pulses depends on the value $V_{{\rm mcp}}$ (see Fig.~\ref{Fig7}): for both coated and non-coated MCPs the larger $V_{{\rm mcp}}$ (i.e. the higher the electron velocity) the larger the width of the delay line pulses. This corresponds with our assumption that the electrons exiting the second plate will have a larger velocity when $V_{{\rm mcp}}$ is larger, therefore increasing their spatial extent at the delay lines (see Appendix~\ref{AppendixA} for details).

Increasing $dV$ to a (small) non-zero value has the effect of collimating and accelerating the electron beam which hits the second micro-channel plate, possibly resulting in a smaller width thanks to a better electron beam collimation (see Section~\ref{Section:GenConsideration}). Experimentally, tuning $dV$ up to 25 V-50 V  diminishes the delay line pulse width (see Fig.~\ref{Fig7}) while further increasing $dV$ leads to a slight increase of this quantity.
Since the width is dominated by dispersion in the transmission lines, its decrease with $dV$ can only be of order of several percent, as it is observed experimentally (about $10~\%$).
The optimal situation might be to work with $dV$ of the order of $25~$V. In order to confirm this statement, we study below the spatially-resolved two-dimensional images extracted from our detector and we measure the resolution in configurations identical to that of Fig.~\ref{Fig7}.

%%%%%%%%%%%%%%%%%%%%%%%%%%%%%%%%%%%%%%
\subsection{Spatial Accuracy}

As mentioned previously, the position of a single incoming particle is extracted from the timing signals out of the delay line anode and for this reason it is related to the pulse amplitude and width studied in the previous sections. We now investigate the accuracy to which such a position can be reconstructed and how this quantity varies with the MCP coating and the interplate voltage $dV$.

A measurement of the accuracy with which the position of a single incoming particle is detected can be obtained as follows from the four arrival times $t_{x1}$, $t_{x2}$, $t_{y1}$ and $t_{y2}$, the fluctuations of which limit the detector accuracy. We consider the quantity $D$ defined as:

\begin{eqnarray}
D = (t_{x_1} + t_{x_2}) - (t_{y_1} + t_{y_2}).
\label{Eq:DefD}
\end{eqnarray}

\begin{figure}[ht!]
\includegraphics[width=.5\columnwidth]{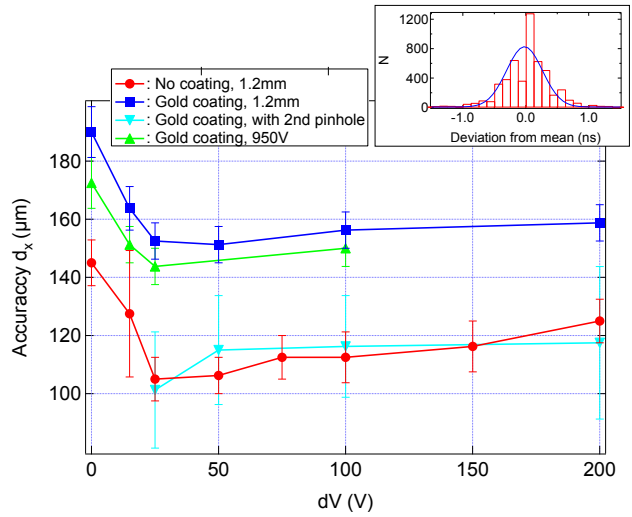}
\caption{\textsl{Spatial accuracy $d_{x}$ (as defined in Eq.~\ref{Eq:SpatialRes}) of the detectors as a function of $dV$ and for different experimental configurations. \emph{Inset:} Typical histogram of the centered distribution of $D$ values measured in the experiment and from which $\sigma_{D}$ is obtained.}}\label{Fig8}
\end{figure}

\noindent Here, $t_{x1}+t_{x2}-2 t_{0}$ and $t_{y1}+t_{y2}-2 t_{0}$ are equal to the propagation time from one end to the other in the two orthogonal delay lines with $t_{0}$ the arrival time of the incoming particle on the delay lines. For identical delay lines, these two quantities are equal and $D=0$. In practice the situation is more complicated. On the one hand, the total lengths of the two delay lines are not strictly equal, leading to an offset for $D$ (we note that in the experiment this offset depends on the actual position we are looking at on the delay line). On the other hand, the quantity $D$ fluctuates from one detected event to the other, leading to a broadening of the distribution of $D$ around its mean value (see inset in Fig.~\ref{Fig8}). We define the accuracy of our detector from the standard deviation $\sigma_D$ of the statistical distribution of $D$. In spatial units, the accuracy is given by (see Appendix~\ref{AppendixB} for details), with $c$ the speed of light:

\begin{equation}
d_x (m) = \frac{\sigma_{D}}{2 \sqrt{2}} \frac{c}{300}.
\label{Eq:SpatialRes}
\end{equation}

In Fig.~\ref{Fig8} we plot the mean spatial accuracy $(d_{x}+d_{y})/2$ as a function of $dV$ and for different configurations. At a fixed flux of incoming particles, we observe that the detector without coating has a better spatial accuracy than the MCP with gold coating. Also, the accuracy of the gold-coated detector is better when the flux is lower. This effect may be related to the changes in the pulse height distribution as a function of the flux and discussed in \cite{Fraser91}: at higher detected flux the pulse height distribution may broaden, leading to a smaller signal-to-noise ratio. We do not observe a great difference between different voltages $V_{{\rm mcp}}$.

Finally we observe that the accuracy of the detector decreases at low voltage values $dV$ and seems optimal for $dV \simeq 25~$V. This optimum $dV$ value for the accuracy corresponds to the optimum values for both the amplitude and the width of the pulses as reported in the previous paragraphs. Nevertheless we believe the pulse amplitude - leading to an improved signal-to-noise ratio -  to have a much larger effect on the accuracy than the pulse width as we shall explain now. Firstly, the decrease in the accuracy (typically $\sim 20 \%$) is much larger than that observed for the width (see Fig.~\ref{Fig7}) while compatible with the increase in the pulse amplitude (see Fig.~\ref{Fig6}). Secondly, in order to identify the underlying mechanism of our observations we have mounted an MCP pair with a voltage difference $dV$ on top of a phosphor screen (instead of a delay line anode). We observed that increasing $dV$ from $0~$V to $\sim100~$V slowly collimates the electron beam, to finally reach excitation of almost a single channel in the second plate. On the other hand, the spot intensity on the phosphor screen strongly increases for $dV$ in the range 0 V-50 V (electrons accelerated in between MCPs having a higher probability of exciting secondary electrons in the second MCP, see Section~\ref{Section:GenConsideration}) to slightly decrease for values of $dV \geq 100~$V. From these complementary studies and our observations on the pulse amplitude and width, we conclude that the main effect for the observed smaller accuracy comes from a better signal-to-noise ratio rather than from the interplate spatial collimation (even though the latter should have a small positive effect as well).

We attribute the latter increase of $d_{x}$ for $dV>50~$V to two origins: \textit{(i)} the signal-to-noise ratio slightly worsens as the pulse amplitude decreases with increasing $dV$; \textit{(ii)} the increase in the electron beam velocity with $dV$ broadens the spot of the electron shower on the delay line anodes (see Appendix~\ref{AppendixA}).
\newline

%%%%%%%%%%%%%%%%%%%%%%%%%%%%%%%%%%%%%%
\section{Conclusion}
In this work we have made a detailed investigation of the effect of a microlayer of gold on the input layer of a MCP and adding a voltage between two slightly separated plates. We find that the detector with the gold layer exhibits larger count rates than the detector without coating, confirming the results of \cite{Tremsin1996}. In addition, pulses extracted from the front of the MCP and pulses obtained from the delay lines show that the gold coated plate yields larger electron showers, leading to a better quantum efficiency. However we demonstrate here that the gold layer has a negative effect on the spatial accuracy, which is crucial for the 3D reconstruction of particles falling onto the detector. At a given flux of incoming particles the detector with the gold coating has an accuracy of $30 \%$ worse than the one without gold. We find that a voltage difference $dV$ between the plates both increases the height of the pulses and reduces the width of the pulses on the delay line, by collimating the electron beam in between the two plates and increasing the energy of the electrons. Obviously, such behavior is very dependent on the distance between the two plates. The addition of $dV$ results in an increase of $~20 \%$ in accuracy for both the coated and non coated plates, an effect coming from an improved signal-to-noise ratio.
Our results demonstrate that using a voltage difference possibly combined with a gold coated plate, leads to improved performance of MCPs used in conjunction with delay line anodes. The improvement in the accuracy with the interplate voltage will probably allow more accurate 3D reconstruction of single particle, metastable atoms detectors, whereas the improvement in the efficiency due to the gold layer could allow easier detection at lower incoming fluxes \cite{Schellekens05,Jeltes07,Hodgman11}.

\begin{acknowledgments}
We thank P. Roncin and R. Sellem for discusssions.
Our collaboration was made been possible thanks to the Triangle de la Physique through the grant DETAMI.
We acknowledge support from the LUMAT Federation, the Triangle de la Physique - contract 2010-062T, the IFRAF Institute, the ANR and the ERC - Grant 267 775 Quantatop.
\end{acknowledgments}
%%%%%%%%%%%%%%%%%%%%%%%%%%%%%%%%%%%%%%
\begin{appendix}

\section{Speed of the electronic pulse along the position axis}
\label{AppendixC}

As sketched in Fig.~\ref{Fig2}, the spatial position $x$ of the incoming particle is encoded perpendicularly to the current propagation in the delay line wires. As a result the effective velocity along the position axis $x$ is smaller than the signal speed, the later being close to light speed in vacuum ($v\simeq c/3$).

This effective velocity is equal to the ratio of the distance between two adjacent loops and the time the signal needs to propagate through one loop. In the case of our Roentdek DLD80, the effective velocity is close to $v_{x}=v/100=c/300$.

\section{Time-of-flight between MCPs and delay line anodes}
\label{AppendixA}
The size of the electron shower hitting the delay line strongly depends on the kinetic energy of the electrons. On the one hand, a larger longitudinal velocity at a fixed transverse velocity implies a smaller time-of-flight between the MCPs and the delay lines, therefore leading to a smaller size of the shower on the delay lines. On the other hand, a larger transverse velocity at a fixed longitudinal one leads to a larger size of the shower on the delay lines.

When electrons are further accelerated while propagating through the MCPs (either by increasing the voltage $V_{{\rm mcp}}$ or $dV$), both the longitudinal and transversal kinetic energy of the electron shower exiting the channel plates increase \cite{Eberhardt79}. The net effect of such an acceleration on the size of the electron shower on the delay lines is not straightforward as there is a competition between a smaller time-of-flight and a larger transverse expansion. The following simple argument corroborates the statement that the spot size actually increases with the kinetic energy of the electrons exiting the MCPs.

We take $V_{g}$ the voltage difference and $d$ the spatial distance between the lower face of the MCP and the delay line. We note $eV_{n}$ and $eV_{t}$ respectively the longitudinal and transverse electron energies at the output of the lower MCP face. From the electrostatic force applied onto the electrons we derive the time of flight between MCPs and the delay lines \cite{Rogers82},
\begin{equation}
t_{tof}=\sqrt{\frac{2 m}{e}} \ \frac{d}{V_{g}} \ \left ( \sqrt{V_{n}+V_{g}} - \sqrt{V_{n}} \right ).
\end{equation}
The transverse spatial expansion of the electron shower on the delay lines is then approximately given by
\begin{equation}
\Delta_{t} = \sqrt{\frac{2 e V_{t}}{m}} \ t_{tof}.
\end{equation}
Assuming that $V_{n}$ and $V_{t}$ increases the same way with the voltage $V_{{\rm mcp}}$ applied on the MCPs, one finds that the larger the voltage $V_{{\rm mcp}}$ the larger the electron shower size $\Delta_{t}$ on the delay lines.

\section{Relating the time accuracy of the delay line to the spatial accuracy on the MCPs}
\label{AppendixB}
Let us define $L$ the length of the wire of one delay line. It takes a time $L/v$ for an electronic pulse to travel from one end to the other at the speed $v$.

The origin of spatial coordinates is set at the center of the delay lines and position is encoded perpendicularly to the wire of the transmission line. As explained in  Appendix~\ref{AppendixC}, the effective velocity $v_{x}$ of pulses along the position axis $x$ is smaller than that ($v$) of propagation in the wire. Writing $t_{0}$ the arrival time on the delay line of the electronic shower from the MCP and $x_{0}$ its spatial coordinate, electronic pulses arrive at the end of the delay line wire at times:
\begin{eqnarray}
t_{x,1}=t_{0}+\frac{L}{2v}-\frac{x_{0}}{v_{x}} \\
t_{x,2}=t_{0}+\frac{L}{2v}+\frac{x_{0}}{v_{x}}.
\end{eqnarray}
Therefore the spatial coordinate $x_{0}$ can simply be extracted from $t_{x,1}$ and $t_{x,2}$ as $x_{0}=(t_{x,2}-t_{x,1})v_{x}/2$.
Assuming uncorrelated errors on the measurement of time $t_{x,1}$ and $t_{x,2}$ and equal standard deviation (namely $\sigma_{t_{x,1}}=\sigma_{t_{x,2}}=\sigma_{t}$), we can write the standard deviation of $x$ as $\sigma_{x}^2=v_{x}^2 \sigma_{t}^2/2$. This quantity $\sigma_{x}$ defines what we refer to as the spatial accuracy of our detector. Under these assumptions for both axes of the delay line, the quantity $D$, defined in Eq.~\ref{Eq:DefD}, has a standard deviation of the form $\sigma_{D}^2=4 \sigma_{t}^2$. Combining the last two equations allows us to relate the error on coordinate $x$ to the distribution of $D$ values as
\begin{equation}
\sigma_{x} = \frac{ \sigma_{D} \ v_{x} }{2\sqrt{2}}=\frac{\sigma_{D}}{2 \sqrt{2}} \frac{c}{300}.
\end{equation}

\end{appendix}
%%%%%%%%%%%%%%%%%%%%%%%%%%%%%%%


\begin{thebibliography}{0}%
\makeatletter
\providecommand \@ifxundefined [1]{%
 \@ifx{#1\undefined}
}%
\providecommand \@ifnum [1]{%
 \ifnum #1\expandafter \@firstoftwo
 \else \expandafter \@secondoftwo
 \fi
}%
\providecommand \@ifx [1]{%
 \ifx #1\expandafter \@firstoftwo
 \else \expandafter \@secondoftwo
 \fi
}%
\providecommand \natexlab [1]{#1}%
\providecommand \enquote  [1]{``#1''}%
\providecommand \bibnamefont  [1]{#1}%
\providecommand \bibfnamefont [1]{#1}%
\providecommand \citenamefont [1]{#1}%
\providecommand \href@noop [0]{\@secondoftwo}%
\providecommand \href [0]{\begingroup \@sanitize@url \@href}%
\providecommand \@href[1]{\@@startlink{#1}\@@href}%
\providecommand \@@href[1]{\endgroup#1\@@endlink}%
\providecommand \@sanitize@url [0]{\catcode `\\12\catcode `\$12\catcode
  `\&12\catcode `\#12\catcode `\^12\catcode `\_12\catcode `\%12\relax}%
\providecommand \@@startlink[1]{}%
\providecommand \@@endlink[0]{}%
\providecommand \url  [0]{\begingroup\@sanitize@url \@url }%
\providecommand \@url [1]{\endgroup\@href {#1}{\urlprefix }}%
\providecommand \urlprefix  [0]{URL }%
\providecommand \Eprint [0]{\href }%
\providecommand \doibase [0]{http://dx.doi.org/}%
\providecommand \selectlanguage [0]{\@gobble}%
\providecommand \bibinfo  [0]{\@secondoftwo}%
\providecommand \bibfield  [0]{\@secondoftwo}%
\providecommand \translation [1]{[#1]}%
\providecommand \BibitemOpen [0]{}%
\providecommand \bibitemStop [0]{}%
\providecommand \bibitemNoStop [0]{.\EOS\space}%
\providecommand \EOS [0]{\spacefactor3000\relax}%
\providecommand \BibitemShut  [1]{\csname bibitem#1\endcsname}%
\let\auto@bib@innerbib\@empty
%</preamble>
\end{thebibliography}%


\begin{thebibliography}{21}
\expandafter\ifx\csname natexlab\endcsname\relax\def\natexlab#1{#1}\fi
\expandafter\ifx\csname bibnamefont\endcsname\relax
  \def\bibnamefont#1{#1}\fi
\expandafter\ifx\csname bibfnamefont\endcsname\relax
  \def\bibfnamefont#1{#1}\fi
\expandafter\ifx\csname citenamefont\endcsname\relax
  \def\citenamefont#1{#1}\fi
\expandafter\ifx\csname url\endcsname\relax
  \def\url#1{\texttt{#1}}\fi
\expandafter\ifx\csname urlprefix\endcsname\relax\def\urlprefix{URL }\fi
\providecommand{\bibinfo}[2]{#2}
\providecommand{\eprint}[2][]{\url{#2}}

\bibitem{examples} J. L. Wiza, Nuclear Instruments and Methods, 141(1-3), 439-601 (1979), G. W. Fraser, nucl. Instr. and Meth. {\bf 221}, 115 (1984).

\bibitem{Barat2000} M. Barat, J. C. Brenot, J. A. Fayeton and Y. J. Picard, Rev. Sci. Instrum. {\bf 71}, 5 (2000).

\bibitem{Vassen12} W. Vassen, C. Cohen-Tannoudji, M. Leduc, D. Boiron, C. Westbrook, A. Truscott, K. Baldwin, G. Birkl, P. Cancio and M. Trippenbach, Rev. Mod. Phys. {\bf 84} 175 (2012)

\bibitem{Jagutzki02} O. Jagutzki, V. Mergel, K. Ullmann-Pfleger, L. Spielberger, U. Spillmann, R. D\"oirner, H. Schmidt-B\"ocking, Nucl. Instrum. Methods Phys. Res. A. {\bf 477}, 244 (2002)

\bibitem{Jeltes07} T. Jeltes, J. M. McNamara, W. Hogervorst, W. Vassen, V. Krachmalnicoff, M. Schellekens, A. Perrin, H. Chang, D. Boiron, A. Aspect and C. I. Westbrook, Nature {\bf 445}, 402-5 (2007)

\bibitem{Schellekens05} M. Schellekens, R. Hoppeler, A. Perrin, J. V. Gomes, D. Boiron, A. Aspect, C. I. Westbrook, Science {\bf 310}, 648 (2005)

\bibitem{Hodgman11} S. S. Hodgman, R. G. Dall, A. G. Manning, K. G. H. Baldwin and A. G. Truscott, Science {\bf 331} 1046-1049 (2011).

\bibitem{Jaskula10} J.-C. Jaskula, M. Bonneau, G. B. Partridge, V. Krachmalnicoff, P. Deuar, K.V. Kheruntsyan, A. Aspect, D. Boiron and C. I. Westbrook, Phys. Rev. Lett. {\bf 105}, 190402 (2010).

\bibitem{Altman04} E. Altman, E. Demler and M.D. Lukin, Phys. Rev. A {\bf 70}, 013603 (2004)

\bibitem{Folling05} S. F\"olling, F. Gerbier, A. Widera, O. Mandel, T. Gericke and I. Bloch, Nature {\bf 434}, 481 (2005)

\bibitem{Greiner05} M. Greiner, C.A. Regal, J.T. Stewart and D.S. Jin, Phys. Rev. Lett. {\bf 94} 110401 (2005)

\bibitem{Rom06}  T. Rom, Th. Best, D. van Oosten, U. Schneide, S. F\"olling, B. Paredes, I. Bloch, Nature {\bf 444}, 733 (2006)

\bibitem{Guarrera08} V. Guarrera, F. Fabbri, L. Fallani, C. Fort, K.M.R. van der Stam, M. Inguscio, Phys. Rev. Lett {\bf 100}, 250403 (2008)

\bibitem{Kitagawa11} T. Kitagawa, A. Aspect, M. Greiner and E. Demler, Phys. Rev. Lett. {\bf 106} 115302 (2011)

\bibitem{Lienard05} See for example E. Li\'enard, M. Herbane, G. Ban, G. Darius, P. Delahaye, D. Durand, X. Fl\'echard, M. Labalme, F. Mauger, A. Mery, O. Naviliat-Cuncic, D. Rodriguez, Nuclear Instruments and Methods in Phys. Res. A {\bf 551}, 375-386 (2005)

\bibitem{Fraser91} G.W. Fraser, M.T. Pain, J.E. Lees and J.F. Pearson, Nuclear Instruments and Methods in Phys. Res. A {\bf 306}, 247-260 (1991)

\bibitem{Tremsin1996} A. S. Tremsin, J. F. Pearson, G. W. Fraser, W. B. Feller, P. White, Nuclear Instruments \& Methods in Physics Re. A \textbf{379}, 139-141 (1996).

\bibitem{Rogers82} D. Rogers, and R. F. Malina, Rev. Sci. Instrum. 53, 1438 (1982)

\bibitem{Firmani82} C. Firmani, E. Ruiz, C. W. Carlson, M. Lampton and F. Paresce, Rev. Sci. Instrum. 53, 570 (1982)

\bibitem{Wiza77} J. L. Wiza, P. R. Henkel, and R. L. Roy, Rev. Sci. Instrum. 48, 1217 (1977)

\bibitem{Paresce75} F. Paresce, Applied Optics {\bf14} (12), 2823-4 (1975)

\bibitem{Eberhardt79} E. H. Eberhardt, Applied Optics {\bf 18} (9), 1418 (1979)

\end{thebibliography}
\end{document}